\documentclass[12pt]{article}
\usepackage[mathscr]{eucal}
\usepackage{bm}
\usepackage{bbold}
\usepackage{caption}
\usepackage{multirow}
\usepackage{amsmath}
\usepackage{textgreek}
\usepackage[greek,english]{babel}
\usepackage{xcolor}
\usepackage{graphicx,subfigure}
\usepackage{empheq}
\usepackage{amsfonts}
\usepackage{amssymb}
\usepackage{dsfont}
\usepackage{ytableau}
\usepackage{overpic}
\usepackage{hyperref} 
\usepackage{makeidx}
\usepackage{empheq}
\usepackage{amsthm}

\newtheorem{thm}{Theorem}
\newtheorem{defi}{Definition}

\newtheorem{rem}{Remark}

\newcommand{\be}{\begin{eqnarray}}
\newcommand{\ee}{\end{eqnarray}}
\newcommand{\bc}{\begin{center}}
\newcommand{\ec}{\end{center}}
\newcommand{\nn}{\nonumber \\}
\newcommand{\lb}{\label}
\newcommand{\pd}{\partial}
\newcommand{\p}[1]{(\ref{#1})}
\newcommand{\vecg}[1]{\mbox{\boldmath $#1$}}

\newcommand{\ga}{\lower.7ex\hbox{$
\;\stackrel{\textstyle>}{\sim}\;$}}
\newcommand{\la}{\lower.7ex\hbox{$
\;\stackrel{\textstyle<}{\sim}\;$}}

\begin{document}

\begin{titlepage}

\vspace*{0.2cm}

\begin{center}

{\LARGE\bf Higher derivatives and ghosts through three centuries}

\vspace{2cm}

{\Large Andrei Smilga}\footnote{Smilga@subatech.in2p3.fr} \\

\vspace{0.5cm}

{\it SUBATECH, Universit\'e de Nantes, 
 4 rue Alfred Kastler, BP 20722, Nantes  44307, France.}

\end{center}

\vspace{0.2cm} \vskip 0.6truecm \nopagebreak

   \begin{abstract}
We outline the history of the research of dynamic systems including higher derivatives in the Lagrangian, The story starts in 1744 when  Euler published a book where the Euler--Lagrange differential equations for a HD system were first written. The  next important landmarks were the papers of Ostrogradsky (1850) and of Pais and Uhlenbeck (1950).  Many articles written in the second half of the last century were influenced by a misconception that was widespread at the time --- it was believed that higher derivatives necessarily bring about a violation of unitarity. 
  We understand today that this is not the case and there are many systems with benign ghosts where HD are present but unitarity is not violated. In this review, we  present and discuss several examples  of such (quantum) mechanical systems and 2D field theories. Then we will briefly discuss higher-derivative gravity theories in 4 dimensions. In that case, the question  whether the ghosts there are benign or not  has a special status due to the absence of a nonperturbative Hamiltonian and does not have a definite answer yet.   However, if one restricts oneself to classical theory, one can argue that the ghosts {\it are} benign (benign classical ghostly system is a system that needs an infinite time to reach a singularity, and it seems to the case both in ordinary and in HD gravity: the cosmic censorship conjecture implies that the singularities are always surrounded by horizons, and it takes an infinite time on the clocks of a distant observer to reach a horizon). 

 Finally, I discuss supersymmetric HD theories: conformal supergravity, where the ghosts are also probably benign, and then an interesting supersymmetric HD gauge theory in 6 dimensions. It involves a dimensionless coupling constant, is renormalizable and rather beautiful.  Unfortunately, it is not unitary --- ghosts are not benign there.
  
\end{abstract}

\end{titlepage}

\section{HD dynamic equations}

\setcounter{equation}0

As was mentioned above, the story starts back in 1744 when Leonhard Euler published a book dedicated to a solution of the {\it brachistochrone} and generalized brachistochrone (the so-called {\it isoperimetric}) problem \cite{Euler}. The conventional brachistochrone problem consists in finding the profile of a curve relating the points $x=0, z=h$ and $x = L, z=0$ such that a stone falling down along this curve did so in a minimal time (see Fg.1).

\begin{figure}
\bc
\includegraphics[width=.6\textwidth]{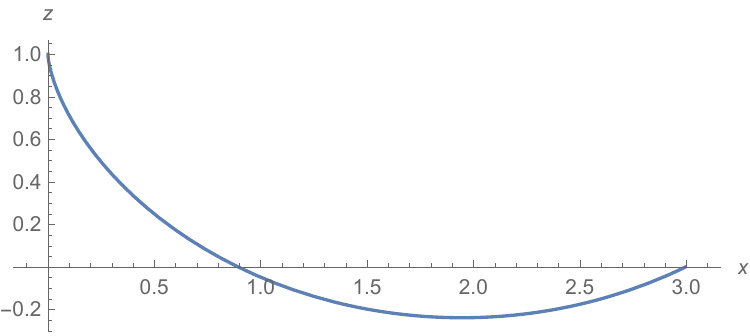}
\ec
        \caption{Brachistochrone, $h=1, L=3$}
\label{brahi}
\end{figure}

 The velocity of the stone at some intermediate point $(x, z)$ is
$$
\frac {ds} {dt} \ =\ \frac {dx\sqrt{1 + (\pd z/\pd x)^2}}{dt} \ =\ \sqrt{2g(h-z)}.
$$
Then the total time of descent is
 \be
\lb{Tfall}
T \ =\ \int dt \ =\ \int_0^L \frac {\sqrt{1 + (\pd z/\pd x)^2}}{\sqrt{2g(h-z)}} dx,
\ee
and the problem is to minimize this integral. As we see, it is a {\it variational} problem. In this particular case, the solution is an arc of a cycloid. This was realised by  Bernoulli brothers, by Leibniz and by Newton at the very end of the 17th century \cite{brachist}. But Euler formuated a general  problem: how find a function $y(x)$ that minimizes the functional
 \be
\lb{I}
I \ =\ \int_a^b Z(x,y,y',...,y^{(n)})\, dx.
\ee
And he wrote the differential equation that the function $y(x)$ must satisfy:
\be
\lb{eqEuler}
\frac {\pd Z}{\pd y} - \frac d{dx} \left( \frac {\pd Z}{\pd y'}\right) + \cdots  + (-1)^n  \frac {d^nx}{dx^n}  \left( \frac {\pd Z}{\pd y^{n}}\right)  \ =\ 0.
\ee
As we now understand well (and Lagrange was the first to notice this), if we interprete $x$ as time, the integral \p{I} acquires the meaning of the {\it action} of a certain 1-dimensional dynamical system, and equation \p{eqEuler} is nothing but what is called now an Euler--Lagrange equation. Generalization to the systems with several degrees of freedom $y_k(x)$ is obvious.

\section{Ostrogradsky's Hamiltonian and its ghostly nature}
\setcounter{equation}0

Mikhail Ostrogradsky was born in 1801 on his family estate in a small Ukranian village {\it Pashennaya}. In 1816--1820 he studied mathematics at the University of Kharkov. Then he moved to Paris and continued his studies in Sorbonna and College de France.
From 1823, he taught at the Lyc\'ee Henri-IV in Paris. He obtained  his most known scientific result during the Parisian period of his life.  In 1826 he  proved what is called now {\it Gauss--Ostrogradsky theorem}  relating the flux of a vector field $\bf{V}$ through a closed surface and the volume integral of the divergence $\bf{\nabla \cdot V} $ inside the surface. He presented it as a report to Paris Academy of Sciences by the title {\it D\'emonstration d’un th\'eor\`eme du calcul int\'egral}  where the manuscipt is available in the academy archive. It was later published in Russian \cite{Ostrolater}. In 1831, he published an abridged version of the proof in \cite{Ostroflux}. Gauss independently proved some special versions of this theorem in his papers of 1813, 1833 and 1839.\footnote{See a  comprehensive historic review in \cite{Katz}.}  

In 1828, he came back to Russia, to St. Petersburg and, in 1830, he was elected a member of Russian Academy of Sciences. He was later elected a corresponding member of many foreign academies.

In 1850, he published a paper reporting his another major well-known scientific result \cite{OstroHam}. Being unaware of Hamilton's papers (they were written in English, which had not at that time the status of of international science language and which Ostrogradsky did not master), Ostrogradsky independently developped the canonical Hamilton formalism and applied it to HD systems.\footnote{It is interesting that, though he talks about {\large Dynamics} in his paper referring to {\large Lagrange} (Ostrogradsky's capslock), a reference to physical dynamical systems is absent in the paper title. Instead, the {\it isoperimetric problem} is mentioned.  Ostrogradsky was, above all, a mathematician and thought like a mathematician. In the XIX century, the rift between pure mathematics and theoretical physics, which is rather deep now, already started to form.} 

Speaking in modern terms, Ostrogradsky traded the first several derivatives in a HD Lagrangian for new dynamical variables and then performed the Legendre transformation. Let us spell it out in the simplest nontrivial case of  a Lagrangian $L(x, \dot x, \ddot x)$, depending only on a single dynamical variable $x(t)$ and its single and double derivatives. The Euler--Lagrange equations of motion read
\be
\lb{eqmot}
\frac {d^2}{dt^2} \frac {\pd L}{\pd \ddot x} - \frac {d}{dt} \frac {\pd L}{\pd \dot x} + \frac {\pd L}{\pd  x}   \ =\ 0.
\ee
This Lagrangian does not depend explicitly on time which entails the energy conservation.\footnote{Evidently, Ostrogradsky was unaware of Noether's theorem, but used it implicitly.} Indeed, one can check that the quantity
\be
\lb{Energy}
E \ =\ \ddot x \frac {\pd L}{\pd \ddot x} + \dot x \left(\frac {\pd L}{\pd \dot x} -  \frac {d}{dt} \frac {\pd L}{\pd \ddot x}  \right) - L
\ee
is an integral of motion.

To construct the Hamiltonian, treat 
\be
\lb{vxdot}
v \ =\ \dot x
\ee
as an independent variable, define the canonical momenta:
\be
\lb{canmom}
p_v  = \frac {\pd L}{\pd \dot v} = \frac {\pd L}{\pd \ddot x} ,  \qquad p_x = \frac {\pd L}{\pd \dot x} - \dot p_v =
\frac {\pd L}{\pd \dot x} -  \frac {d}{dt} \frac {\pd L}{\pd \ddot x},
\ee
express the generalized velocities $\dot x, \dot v$ via $p_x$ and $p_v$ and rewrite the energy \p{Energy} as
\be 
\lb{OstroHam}
H(p_v, p_x;\,  v,x) = p_v \dot v  + p_x \dot x  - L \ = \ p_v a(p_v, x, v) + p_x v - L[a(p_v, x, v), x, v],
\ee
where $a(p_v, x, v)$ is the solution of the equation $\pd L(x; v; a)/\pd a = p_v$. The Hamilton
equations of motion are\footnote{See Eq. (14) in Ostrogradsky's paper.}
 \be
(I)  \ && \frac {\pd H}{\pd p_v} = \dot v, \qquad (II) \  \quad \frac {\pd H}{\pd v} = - \dot p_v \nn
(III)  \ &&\frac {\pd H}{\pd p_x} = \dot x, \qquad (IV)  \ \quad \frac {\pd H}{\pd x} =  - \dot p_x.
 \ee
We see that equation (III) coincides with \p{vxdot}, equations (I) and (II) are fulfilled,
bearing in mind the definition  of the canonical momenta in \p{canmom}, and equation (IV)
is equivalent, bearing in mind all these definitions, to the Euler--Lagrange equation of
motion \p{eqmot}.

If the Lagrangian depends on $x,  \dot x, \ldots,  x^{(n)}$ in a nontrivial way, one should define
$n-1$ new variables:
$$
y_1 = \dot x, \  y_2 = \ddot x, \ldots,  \ y^{(n-1)}  = x^{(n-1)}
$$ 
and the canonical momenta
\be
\lb{canmom-n}
p_{n-1} \ & = &\  \frac {\pd L}{\pd \dot y_{n-1}}, \nn
p_{n-2} \ &=&\  \frac {\pd L}{\pd \dot y_{n-2}} - \dot p_{n-1} \nn
\cdots \ &=& \ \cdots  \nn
p_0\ &=&\  \frac {\pd L}{\pd \dot x} - \dot p_1. 
\ee
Then the Hamiltonian, obtained by the Legendre transformation,
\be
\lb{OstroHam-n}
H \ =\ p_{n-1} \dot y_{n-1} + \cdots + p_0 \dot x - L,
\ee
 involves $n$ pairs of canonically conjugated variables
$(p_0, x),  \ldots  (p_{n-1}, y_{n-1})$. 
Again, the Hamilton equations of motion following from \p{OstroHam-n} are equivalent to the Euler--Lagrange equations of motion following from $L(x, \dot x, \ldots, x^{(n)})$.

The Hamiltonian \p{OstroHam-n} has a remarkable property:

\begin{thm}
\lb{neogran}
 If $n \geq 2$ and the canonical momenta $p_1, \ldots, p_{n-1}$ do not all vanish, the
function \p{OstroHam-n} may acquire an arbitrary real value.
 \end{thm}

This property was not noticed by Ostrogradsky himself. As was mentioned, he was a mathematician and was not interested so much in applications of the formalism that he developped for particular dynamical systems. It was first observed in several particular examples (we will discuss them in the next section) and a general proof was given not so long time ago by Eliezer and Woodard\footnote{See also Theorem 1 in Ref. \cite{Pagani}. These authors proved that  Ostrogradsky's Hamiltonian of a higher-derivative system has no definite sign.} \cite{Eliezer}. Actually, the proof is not complicated.

\begin{proof}

Rewrite \p{OstroHam-n} as 
\be
\lb{OstroHam1-n}
H \ =\ p_{n-1} \dot y_{n-1} + p_{n-2} y_{n-1}  + \cdots + p_0 y_1 - L(\dot y_{n-1}, y_1, \ldots, y_{n-1}, x),
\ee
The generalized velocity $\dot y_{n-1}$ can be expressed via the variables $p_{n-1}, y_1, \ldots, y_{n-1}, x$
by solving the first equation in \p{canmom-n} for $\dot y_{n-1}$. It does not depend on other
generalized momenta $p_0, \ldots,  p_{n-2}$. Thus, the only dependence of the Ostrogradsky's
Hamiltonian  on these momenta is due to the contribution
$$
X \ =\ p_{n-2} y_{n-1}  + \cdots + p_0 y_1.
$$
This contribution is linear in momenta and can obviously acquire any real value.
And the same is true for \p{OstroHam1-n}.

\begin{rem}
{\em The requirement $n \geq 2$ is essential. Ordinary dynamical systems with
$n = 1$ include only one generalized momentum $p_0 \equiv p_{n-1}$, and the dependence of
the Hamiltonian on $p_0$ need not be linear. For example, the familiar Hamiltonian
$H = p^2/2$ for a free particle is quadratic in $p$ and takes only positive values.}
\end{rem}

\begin{rem}
{\em All this can be easily generalized on the case of several variables $x_j$.
The theorem above holds if the dependence of the Lagrangian on at least one of
the higher derivatives of some such variable is nontrivial so that the corresponding
partial derivative of $L$ does not vanish.}
\end{rem}

\end{proof}

In 2015, Richard Woodard published an influential review \cite{Woodard} which starts with the detailed exposition of the Ostrogradskian formalism and of the proof  above. The exposition and the proof are good, but the {\it title} of his paper, ``Ostrogradsky's theorem on the Hamiltonian instability", is not in my personal opinion quite well chosen. First of all, this theorem belongs not to Ostrogradsky but rather to Pagani, Techhiolli and  Zerbini (in a restricted version) and to Eliezer and Woodard (in the full version). Second, the word {\it instability} may bring (and it {\it does} often bring) wrong associations. We know well the Lyapunov instability associated with chaotic behavior:  a small change of the initial conditions at $t=0$ leads to large deviations at later times.  Here we are dealing with an irrelated and  completely different phenomenon:  a classical HD system admits arbitrarily high and arbitrarily low energies. As we will see soon, this feature persists also in quantum theory: the ground state in the spectrum (the vacuum) is absent.

To establish our terminology, we give here a couple of formal definitions.

\begin{defi}
A  classical ghostly system is a system whose classical Hamiltonian is not bounded neither from below nor from above.
\end{defi}

\begin{defi}
A  quantum ghostly system is  a system with the spectrum not bounded neither from below nor from above.
 \end{defi}
And the word {\it instability} will always mean  Lyapunov instability. 

Not only HD systems have ghostly nature. The systems including highly attractive potentials also exhibit this feature --- think e.g. of 3D motion in the  potential $V(r) = -\gamma/r^2$.

Quantum ghosts are usually associated with  loss of unitarity  (like the Faddeev--Popov ghosts, which may be present in the loops but not as asymptotic states). In some quantum ghostly systems, the existence of ghosts  may lead to violation of unitarity, indeed. But it is not necessary. We will discuss this at length later, and for the time being we present two more definitions. 

\begin{defi}
\lb{ben-clas-ghost}
A {\em benign} classical ghostly system is a classical ghostly system not plagued by a blowup (alias collapse) and enjoying classical trajectories that are regular at all times.  
\end{defi}

\begin{defi}
A  benign quantum ghostly system is a quantum ghostly system with a unitary evolution operator.
 \end{defi}

\section{Pais--Uhlenbeck oscillator}
\setcounter{equation}0

During a century since 1850,  HD systems (mostly, HD field theories) were sporadically discussed. (See e.g. the review \cite{Podolsky} dedicated to a HD electrodynamics.) But an important progress in understanding the classical and quantum HD dynamics is associated with the 1950 paper of Pais and Uhlenbeck \cite{PU}.

The {\it Pais--Uhlenbeck oscillator} is described by the following HD Lagrangian:
\be
\lb{L-PU}
L \ =\ \frac 12 \left[\ddot x^2  -  (\omega_1^2 + \omega_2^2) \dot x^2  + \omega_1^2 \omega_2^2 x^2 \right],
\ee
The corresponding Ostrogradsky's Hamiltonian reads
\be
\lb{H-PU}
H \ =\ p_x v + \frac {p_v^2}2 + \frac {(\omega_1^2 + \omega_2^2)v^2}2 - \frac{\omega_1^2 \omega_2^2 x^2 }2.
\ee
One of the Hamilton equations of motion gives $\pd H/ \pd p_x = v = \dot x$. 

The quantum Hamiltonian $\hat H$ also has the form \p{H-PU} with $\hat p_x = -i\pd/\pd x, \hat p_v = -i\pd/\pd v.$ Consider first the case of inequal frequencies  and assume for definiteness that $\omega_1 > \omega_2$. To determine the  spectrum of $\hat H$, it is convenient to perform the following quantum canonical
transformation \cite{Pagani,Mannheim}:
\be
\lb{can-trans}
&&X_1   = \frac 1{\omega_1} \frac {\hat p_x + \omega_1^2 v}{\sqrt{\omega_1^2 - \omega_2^2}}, \qquad \hat P_1 = -i \frac \pd{\pd X_1} = \omega_1  \frac {\hat p_v + \omega_2^2 x}{\sqrt{\omega_1^2 - \omega_2^2}}, \nn
&&X_2   =  \frac {\hat p_v + \omega_1^2 x}{\sqrt{\omega_1^2 - \omega_2^2}}, \qquad \qquad \hat P_2 = -i \frac \pd{\pd X_2}  =
 \frac {\hat p_x + \omega_2^2 v}{\sqrt{\omega_1^2 - \omega_2^2}}.
\ee
In these terms, the Hamiltonian reduces to a difference of two oscillator
Hamiltonians:
\be
\lb{two-osc}
\hat H \ =\ \frac {\hat P_1^2 + \omega_1^2 X_1^2}2 - \frac {\hat P_2^2 + \omega_2^2 X_2^2}2
\ee
Its spectrum is
\be
\lb{PU-spectrum}
E_{nm} \ =\ \left(n + \frac 12  \right) \omega_1 - \left(m + \frac 12  \right) \omega_2
\ee
with positive integer $n,m$. 

One may ask: how come the spectrum is asymmetric under interchange $\omega_1 \leftrightarrow \omega_2$,
whereas the Lagrangian \p{L-PU} and the Hamiltonian \p{H-PU} have this symmetry? The asymmetry was brought about by the assumption $\omega_1 > \omega_2$. In the opposite case, the canonical transformations \p{can-trans} should be modified to include $\sqrt{\omega_2^2 - \omega_1^2}$ rather than $\sqrt{\omega_1^2 - \omega_2^2}$,
 and the sign of the energy
in \p{PU-spectrum} would  then be reversed.

The classical motion of the system \p{L-PU}  with inequal frequencies is finite as for an ordinary oscillator. The quantum dynamics also does not have a lot of  pecularities. If the frequencies $\omega_1$ and $\omega_2$ are commensurable (say, $\omega_1 = 2\omega_2$), the spectrum \p{PU-spectrum} is discrete with an infinite degeneracy on each level. If the ratio $\omega_1/\omega_2$ is irrational, the spectrum is still pure point but everywhere dense.  The evolution operator for the system is simply a product of the evolution operators for the individual oscillators:
  \be
\lb{evol-oper}
 {\cal K}(X_1', X_2'; X_1, X_2; t ) \ =\ \sum_{n=0}^\infty \psi^*_n(X_1') \psi_n(X_1) \exp\left\{-i\omega_1(n + 1/2)t  \right\} \nn
\times  \sum_{m=0}^\infty \psi^*_m(X_2') \psi_m(X_2) \exp\left\{i\omega_2(m + 1/2)t  \right\},
 \ee
where  $\psi_n(X)$ are the standard oscillator eigenfunctions. The system is unitary. 

Finally, if $\omega_1 = \omega_2$, the canonical transformation \p{can-trans} is impossible and the situation is slightly more complicated. 
To begin with, the motion is not finite anymore. Besides ordinary oscillations 
$x(t) \sim e^{i\omega t}$, the oscillations with linearly rising amplitude, $x(t) \sim t e^{i\omega t}$ appear. But the presence of such mildly rising classical solutions does not destroy quantum unitarity: the Hilbert space  with a well-defined norm and inner product is still well defined for the corresponding Schr\"odinger problem. 

One can show \cite{PU,Bolonek,Singapur} that in this case the spectrum is continuous. The explicit form of the eigenfunctions of the Hamiltonian
\be
\lb{H-free-degen} 
\hat H  \ =\ -iv\frac \pd{\pd x} - \frac 12 \frac {\pd^2}{\pd v^2} + \omega^2 v^2 - \frac {\omega^4 x^2}2
\ee
is
\be
\lb{Psi-degen}
\Psi_{nk}(x,v; t) \ =\ e^{-i\omega xv} \exp\left\{ \frac i{4\omega^2} \frac {\pd^2}{\pd x \pd v} \right\} \left[J_n(k \sqrt{v^2 + \omega^2x^2}) \left( \frac{\omega - iv}{\omega + iv} \right)^{n/2}  \right] e^{-it(n\omega + k^2/4)}
 \ee
with $n = 0, \pm 1, \pm 2, \ldots$ and $k \in {\mathbb R}$.

Any real energy $E_{nk}  = n\omega + k^2/4$ is admissible and each energy level is infinitely degenerate.

Thus, the PU oscillator is a benign ghostly system both in the classical and in the quantum sense.

Following Ref. \cite{PU}, we have discussed up to now only the PU oscillator, the simplest nontrivial HD system. What happens in more complicated case when the Lagrangian includes nontrivial interactions and the Schr\"odinger problem cannot be exactly solved?

One can prove  in this case a general theorem \cite{eston}, a quantum counterpart of the classical Theorem \ref{neogran}.
\begin{thm}
The quantum counterpart of the Ostrogradsky Hamiltonian for a non-degenerate higher-derivative system has no ground state.
\end{thm}
\begin{proof}
Following Ref. \cite{Singapur}, we will outline here the proof of this fact in the simplest case of the system $L(x, \dot x, \ddot x)$ 
with $\pd L/\pd \ddot x  \neq 0$. We will also assume that the spectrum is discrete.

Introduce as earlier $v = \dot x$ and consider the quantum Hamiltonian
\be
\hat H\ =\
a(\hat p_v, x, v)\hat p_v + v \hat p_x  -  L[a(\hat p_v, x, v), x,  v],
 \ee 
 where $a$ is the solution of the equation $\pd L(x, v;\, a)/\pd a = p_v$.
Its middle term is $-iv\pd/\pd x$. Suppose that this Hamiltonian has a normalized ground
state 	$\psi_0(x, v)$ of energy $E_0$. We take a variational ansatz
\be
	\psi_c(x, v) \ =\  e^{icx}\psi_0(x, v).
\ee 
Obviously, 
$$
	\int \psi^*_c \psi_c \, dx dv \ =\ \int \psi^*_0 \psi_0 \,  dx dv \ =\ 1.
$$ 
The function $\psi_c(x, v)$  is not an eigenfunction of the Hamiltonian:
  $$
\hat H \psi_c  = E_0	\psi_c + c v	\psi_c \neq \lambda \psi_c 
  $$
Then $\psi_c$ should have projections on the excited states in the spectrum and the variational energy 
$E(c) =  \int \psi^*_c \hat H \psi_c\, dx dv  $ should exceed $E_0$. However, the direct calculation gives:
\be
\lb{Ec}
E(c) \ =\ E_0 + c \int v |\psi_0|^2 \, dx dv.
 \ee
If the integral $\int v |\psi_0|^2  \,dx dv$ is different from zero, we can shift the energy down
by choosing an appropriate sign of $c$. If this integral is zero, the energy does not
depend on c and does not grow, as it should if $\psi_0$ were  a ground state. Thus,
the original assumption of the existence of the ground state was wrong.
\end{proof}
In other words, {\it a classical ghostly system corresponds to a quantum ghostly system}.

\vspace{1mm}
An important question that can be posed to a particular quantum HD theory is whether the ghosts there are {\it benign} (i.e. in spite of its unusual structure, the absence of a ground state, the theory is unitary with a  well-defined evolution operator) or {\it malignant}
and unitarity is violated.  

Well, it depends.

\section{Ghosts and unitarity}
\setcounter{equation}0

Recall the standard way to treat a quantum field theory. Based on the kinetic part of the Lagrangian, one defines the creation and annihilation operators $a_k^\dagger$ and $a_k$. The vacuum state $|0\rangle$ is defined as the state annihilated by the action of $a_k$. The states $a_k^\dagger |0\rangle$ are one-particle states. Then one defines the $S$-matrix including the transition amplitudes between  the one-particle asymptotic states. This works well for ordinary theories with conventional positive kinetic terms.

Let the theory includes only bosonic fields\footnote{Including fermionic fields provides some technical modifications but does not change the essense.} with ordinary quadratic  terms,
$$
 {\mathcal L}^{(2)} \  = \ \frac 12 \sum_i  \left[ (\pd_\mu \phi_i)^2 -  m_i^2 \phi_i^2 \right].    
$$
The corresponding quantum Hamiltonian in a finite box, normalized such that the vacuum state has zero energy, reads
 \be
\hat H^{(2)}  \ =\ \sum_i \sum_{\mathbf{k}_i} a_{i\mathbf{k}_i}  a^\dagger_{i\mathbf{k}_i} \sqrt{m_i^2 + \mathbf{k}_i^2}.
\ee
The creation and annihilation operators satisfy the commutation relations 
 \be
[a_{i\mathbf{k}}, a^\dagger_{j\mathbf{k}'}] \ =\ \delta_{ij} \delta_{\mathbf{k}\mathbf{k}'}.
\ee
Then the excited states like $a^\dagger_{i\mathbf{k}}|0\rangle$ have positive energy and positive norm.

Consider now a theory including two fields $\phi_{1,2}$ with a non-standard kinetic term,
 \be
\lb{plus-minus}
T \ =\ \frac 12 \left[ (\pd_\mu \phi_1)^2 -  (\pd_\mu \phi_2)^2  \right].
\ee
The quadratic Hamiltonian is now
 \be
\lb{Ham-quadr}
\hat H \ =\  \sum_{\mathbf{k}} a_{1\mathbf{k}}  a^\dagger_{1\mathbf{k}} \sqrt{m_1^2 + \mathbf{k}^2} 
-  \sum_{\mathbf{k}} a_{2\mathbf{k}}  a^\dagger_{2\mathbf{k}} \sqrt{m_2^2 + \mathbf{k}^2} 
  \ee
The energy can be both positive and negative. This means that the state $|0\rangle$ defined as earlier such that $a_{1,2} |0\rangle = 0$ is not the state with the lowest energy anymore. The vacuum is absent.

Can one somehow eliminate this unpleasant drawback and define on the base of the Lagrangian \p{plus-minus} a Hamiltonian whose spectrum would have a vacuum? Yes, one can.  It is sufficient to take the same Hamiltonian \p{Ham-quadr} but redefine  the vacuum state. We require the new vacuum $|\Omega \rangle$ to satisfy the conditions 
\be
\lb{newvac}
a_{1\mathbf{k}} |\Omega\rangle =0, \qquad {\rm and} \qquad a_{2\mathbf{k}}^\dagger |\Omega\rangle =0.
 \ee 
Also we assume  that $\langle \Omega | \Omega \rangle =1$. In fact, this assumption is not so innocent. For a  2-dimensional oscillator system with the Hamiltonian $H = a_1 a_1^\dagger - a_2 a_2^\dagger$, the wave function $|\Omega \rangle$ satisfying the conditions \p{newvac} reads $|\Omega\rangle \propto e^{-X_1^2/2  } e^{X_2^2/2  }$. Such $|\Omega \rangle$ is not normalizable in the ordinary sense and is pretty much unphysical.

Not to clutter up the formulas, we stay with a 2-dimensional system. The state $|{\large 1} \rangle  = a_1^\dagger |\Omega \rangle$ is an ordinary state with positive norm, $\langle {\large 1}| {\large 1} \rangle = 1$. Consider, however, the state\footnote{$a_2^\dagger |\Omega \rangle =0$, as we remember.}   $|{\large 2}\rangle  = a_2 \Omega$. It has a {\it negative} norm:
$$
\langle {\large 2}| {\large 2} \rangle  \ =\ \langle \Omega | a_2^\dagger a_2 |\Omega\rangle \ =\ -1,
$$
where the relation $[a_2, a_2^\dagger] = 1$ was used. 

On the other hand, the energy of both excited states $|{\large 1}\rangle$ and $|{\large 2} \rangle$ has the same positive value. We have,
$$ E_1 \ =\ \frac {\langle {\large 1}|\hat H| {\large 1} \rangle}{\langle {\large 1}|{\large 1} \rangle} \ =\ \langle \Omega|a_1(a_1 a_1^\dagger - a_2 a_2^\dagger) a_1^\dagger|\Omega \rangle \ =\ 2 $$
and also\footnote{For any state $\Psi$ with positive or negative norm, its energy $E_\Psi$ is defined according to $\hat H \Psi   = E_\Psi \Psi$. Hence,
$$
E_\Psi  =\ \frac{\langle \Psi|\hat H|\Psi \rangle}{\langle \Psi|\Psi \rangle}.  
$$}
$$ E_2 \ =\ \frac {\langle {\large 2}|\hat H| {\large 2} \rangle}{\langle {\large 2}| {\large 2} \rangle} \ =\ -\langle \Omega|a_2^\dagger (a_1 a_1^\dagger - a_2 a_2^\dagger) a_2 \rangle | \Omega\rangle \ =\ 2 $$

It is nice to have positive energies, but appearance of the states with negative norm represents a problem.  Creation of such states  means violation of  unitarity. The cure happened to be worse than the desease \cite{Kovner}.

If we travel back in time for 20 years or more, find any quantum-field  expert and ask their opinion about higher-derivative theories, the answer would be that such theories are poisoned by the presence of negative-metric states and associated unitarity violation and are not acceptable by that reason. Well, there were some attempts to find an antidote. The most known such attempted antidote was the {\it Lee--Wick mechanism}  \cite{LeeWick}. These authors discussed a modified electrodynamics including besides the ordinary photon also a massive boson with the negative residue of the pole of   propagator, such pole being  associated with the negative metric\footnote{They did so to handle  ultraviolet divergences  like in the Pauli-Villars regularization scheme.}. It was observed that this negative residue pole  acquires an imaginary part after taking the perturbative corrections  into account and do not correspond to an asymptotic state anymore. However, it was simultaneously observed that such theory looses causality. 

If you stick to the requirement that the theory should have a ground vacuum state in the spectrum, there is no satisfactory remedy, indeed.\footnote{Well, one of the popular suggestions to ``exorcise" the ghost while keeping the energy positive was to take the system \p{two-osc} and {\it complexify} the variable $X_2$ (and correspondingly $P_2$) \cite{Bender}. Then the second term in the Hamiltonian changes sign, making the energy positive definite. But this means {\it redefining} the spectral Sturm-Liouville problem. The energy becomes positive and the ghost is busted in this redefined problem but not in the original one.}. Higher-derivatives lead to bad consequencies. And it was only comparatively recently understood that the problems disappear if one abandons this strict requirement.

\section{Benign ghosts}

\setcounter{equation}0

Of course, people always understood that a {\it free} ghostly system with the Hamiltonian like in Eq. \p{two-osc}  is quite consistent and admissible. After all, an exact solution to the Schr\"odinger equation exists in this case. There are stationary states, with some of them having negative energies going down to $-\infty$. Unusual but admissible. But a common lore used to be that, as soon as interactions are switched on, the system ceases to be stationary. The energy would flow from the positive energy sector to the negative energy sector in an unruly way leading to collapse.

Indeed, this is what happens in many cases. But not in {\it all} cases.

To the best of my knowledge, the first paper where an interactive  ghostly QM system was considered  was Ref. \cite{Pagani}. These authors considered the following Hamiltonian:
 \be
\lb{L-Pagani}
H \ =\  \frac {\hat P_1^2 + \omega_1^2 X_1^2}2 - \frac {\hat P_2^2 + \omega_2^2 X_2^2}2 + V(X_1 - X_2)
 \ee
with an irrational ratio $\omega_1/\omega_2$ and interaction chosen in the form
 \be
V(q) \ =\ \mu_3 q^3 + \mu_4 q^4 + O(q^5).
 \ee 
with $\mu_3$ and $\mu_4$ not vanishing simultaneously. Under these condition, they proved a {\it theorem}  that such system is stable in the Lyapunov sense in a certain neighborhood of the equilibrium point $X_1 = P_1 = X_2 = P_2 = 0$.

Nothing happened in this direction for some time afterwards, but in 2003 the paper \cite{Carroll} appeared.   Its authors studied the following model Hamiltonian\footnote{They were interested in it, bearing in mind some cosmological implications.}:
 \be
\lb{H-Carroll}
H \ =\ \frac 12 [p_x^2 - (p_y^2 + y^2) + x^2 y^2],
  \ee
where a free particle is coupled to the ghost oscillator. If we pose the initial conditions
$$
x(0) = y(0) = a, \quad  p_x(0) = p_y(0) = 0, 
$$
the system undergoes benign oscillations at the vicinity of the origin if $a \leq  0.59$ 
and goes astray, running to infinity, for larger deviations. The system thus includes an {\it island of stability}.
A similar island of stability was later found \cite{benign} for the Lagrangian of the PU oscillator complemented by a quartic term:\footnote{We dwell here on the system with degenerate frequencies, but the physics is basically the same also when $\omega_1 \neq \omega_2$.} 
\be
\lb{PU+x4}
L \ =\ \frac 12 
\left[  \ddot x^2 - 2\omega^2 \dot x^2 + \omega^4 x^2 \right] - \frac 14 \alpha x^4.
\ee

\begin{figure}[ht!]
   \begin{center}
 \includegraphics[width=4 in]{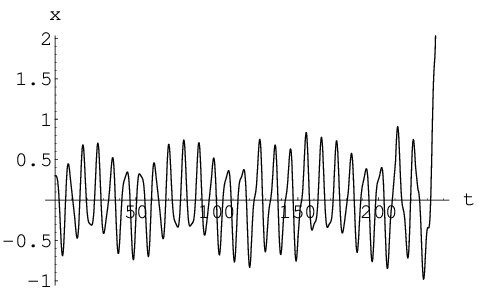}
    \end{center}
\caption{Oscillating and collapsing}
\label{island}
\end{figure} 

For positive $\alpha$, the system is stable at the vicinity of the point $x = \dot x = \ddot x = x^{(3)} = 0$. Let us slightly deviate from this point and impose the initial conditions
 \be
x(0) = c, \quad \dot x(0) = \ddot x(0) = x^{(3)}(0) =0.
  \ee
It occurs that, for small enough $c < c_{\rm crit}  \approx .3\omega^2/\sqrt{\alpha}$, the system exhibits benign oscillations around zero. And for larger $c$ we are beyond the island of stability and the system goes astray running to infinity. (See Fig. \ref{island} where the behavior of the system slightly above $c_{\rm crit}$ is displayed.)  

In the systems \p{H-Carroll} and \p{PU+x4}, the ghosts are not completely benign: when the initial conditions are outside the island of stability, a blowup occurs. 
The first example of a system with completely benign ghosts was found in Ref. \cite{Robert}. It is a mechanical system with the Lagrangian\footnote{Genetically, it comes  from  in a certain supersymmetric (quantum) mechanical systems but we will not distract here from our main line of discussion. The details may be found in the original paper \cite{Robert} and in the review \cite{Singapur}.} 
\be
\lb{L-Robert}
L \ =\ \dot x \dot D - DV'(x).
\ee
 The corresponding Hamiltonian
involves two pairs of dynamic variables, $(x, p)$ and $(D, P)$.  It reads
  \be
\lb{H-Robert}
H  = \ pP + DV'(x),
 \ee
where $V (x)$ is an arbitrary smooth even function. The simplest nontrivial case is
$$
 V (x) =  \frac {\omega^2 x^2}2 + \frac {\lambda x^4}4.
 $$
 \begin{figure}[h]
   \begin{center}
 \includegraphics[width=2.4 in]{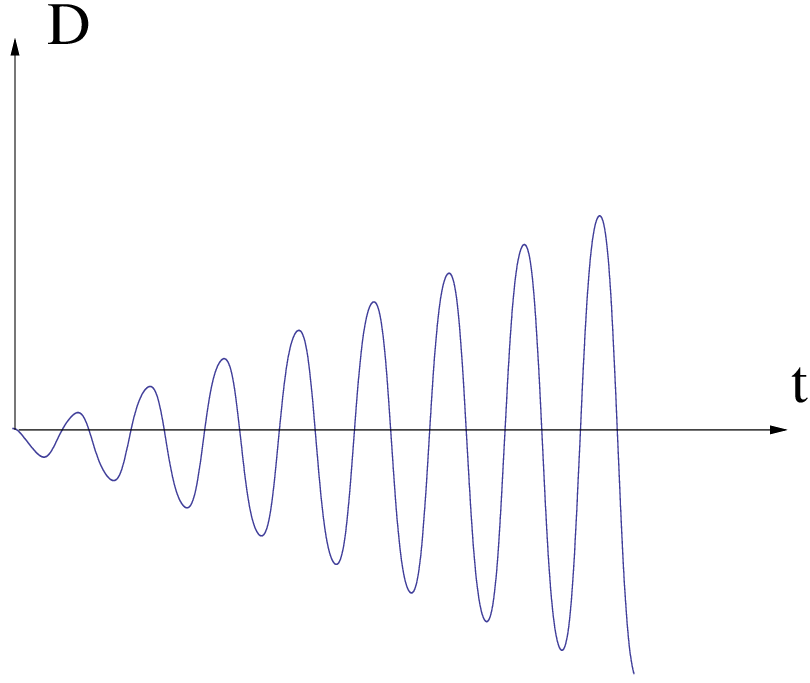}
    \end{center}
\caption{A typical behaviour of $D(t)$.} 
\label{Drost}
\end{figure} 

The kinetic part of the Hamiltonian \p{H-Robert} is not positive definite, and, obviously,
the spectrum of its quantum counterpart has no bottom. If one introduces 
$$
X_{1,2} \ =\ \sqrt{\frac \omega 2}x  \pm \frac 1{\sqrt{2\omega}}D
$$
and the corresponding canonical momenta, the Hamiltonian acquires the form
 \be
H \ =\ \frac {P_1^2 + \omega^2 X_1^2}2 -  \frac {P_2^2 + \omega^2 X_2^2}2 + \frac \lambda{4\omega}(X_1-X_2)(X_1 + X_2)^3.
\ee
In other words, this is the Pais-Uhlenbeck Hamiltonian \p{two-osc} with degenerate frequencies, where an
extra quartic interaction of a special form is added.

A nice distinguishing feature of the system \p{H-Robert} is its exact solvability. Indeed,
it involves besides $H$ another integral of motion\footnote{The kinetic part of the Hamiltonian \p{H-Robert} is not positive definite, and that is why this system was missed in the classification of exactly soluble Hamiltonian systems with two degrees of freedom in the book \cite{Perelomov}.}
\be
\lb{N}
N  = \frac {P^2}2 + V (x).
 \ee
The Poisson bracket $\{N, H\}$ vanishes. This allows one to find the solution analytically. 
It is expressed via certain elliptic functions. $x(t)$ is simply an elliptic cosine with the period depending on $\omega, \lambda$ and the integral of motion $N$. The function $D(t)$ also exhibits an oscillatory behavior, but with a linearly rising amplitude (see Fig. \ref{Drost}).
A linear growth is not a blowup, however. The ghost  is benign here.

\vspace{1mm}

Integrabily of the model \p{H-Robert} allowed one to find exactly  the spectrum and eigenfunctions of the quantum counterpart of the Hamiltonian \cite{Robert}. The pattern is the same as for the free Hamiltonian \p{H-free-degen} --- the spectrum is continuous  running from $-\infty$ to $\infty$, with all the states being infinitely degenerate. 
Unitarity is preserved. Both classical and quantum versions of the  ghostly system \p{H-Robert} are benign. 

But integrability is not at all a necessary condition for the ghosts to be benign. One can for example modify the model \p{H-Robert} and consider the Hamiltonian \cite{Robert}
\be
\lb{H-Robert-gam}
H  = \ pP + DV'(x) - \frac \gamma 2D^2 + P^2).
 \ee
Such a model is not integrable anymore and its analytic solution is not possible. However, this system can be studied numerically.   The trajectories exhibit a chaotic behavior, but do not exhibit any blowup \cite{Robert}, the ghosts are benign.

Later many other benign ghostly systems were found. In Refs. \cite{Pavsic} the benign nature of the classical Hamiltonian \p{two-osc} complemented by the interaction term
 \be
\lb{V-Pavsic}
V(X_1, X_2) = \lambda \sin^4 (X_1+X_2)
 \ee
was demonstrated. The authors of Ref. \cite{Kovner} arrived at the same conclusion for the interaction
\be
\lb{V-Kovner}
V(X_1, X_2) = \lambda (X_1^4 - X_2^4) + \mu X_1^2 X_2^2, \qquad \qquad \lambda \gg \mu.
 \ee
And the authors of Ref. \cite{Deffayet} showed that for the interaction
\be
\lb{V-Deffayet}
V(X_1, X_2) \  = \  \frac \lambda {\sqrt{(X_1^2 - X_2^2 -1)^2 + 4 X_1^2}}.
 \ee

\vspace{1mm}
There are also many  ghostly systems unrelated to the PU oscillator. 
We dwell on two such examples. 
\begin{itemize}

\item In Ref. \cite{Damour}, the benign nature of ghosts in the system describing motion along a geodesically complete Lorentzian manifold, e.g. the AdS manifold with the metric
 \be
\lb{metr-AdS}
ds^2 \ =\ - \left(1 + \frac {r^2}{l^2}   \right) dt^2 + \frac {dr^2}{1 + \frac {r^2}{l^2}} + r^2 d\Omega_{D-2}^2,
 \ee
was established.

\item In Ref. \cite{Toda}, we analyzed the well-known Toda system and found out that it also can be interpreted as a ghostly system.

Consider the Hamiltonian of the closed Toda chain with three particles:
\be
\lb{Toda}
H \ =\ \frac 12(p_1^2 + p_2^2 + p_3^2) + V_{12} + V_{23} + V_{31}
 \ee
with $V_{12} = e^{q_1-q_2}$ etc. This Hamiltonian is positive definite and it seems that there are no ghosts. However,  
besides the energy and the obvious linear integral of motion $P = p_1 + p_2 + p_3$, the system involves also cubic one:
\be
\lb{cubic}
I \ =\ \frac 13(p_1^3 + p_2^3 + p_3^3) + p_1 (V_{12} + V_{31}) + p_2 (V_{12} + V_{23}) + p_3 (V_{23} + V_{31}).
 \ee
The Poisson brackets $\{I, H\}, \, \{P, H\}$ and $\{I,P\}$ vanish. For the system with three degrees of freedom, we have three integrals of motion that are in involution. The system is integrable.

Now, the main idea is to treat the invariant $I$ rather than $H$ as a Hamiltonian. Being cubic in momenta, it is not bounded from below, neither from above. From the viewpoint of $I$, the Toda system is a ghostly system. The new Hamiltonian implies the new Hamiltonian equations of motion. Their explicit form is
 \be
\lb{eqmot-I}
\dot q_1 = p_1^2 + V_{12} + V_{31}, \qquad \qquad \dot p_1 = (p_1 + p_3) V_{31} -(p_1 + p_2) V_{12}, \nn
\dot q_2 = p_2^2 + V_{23} + V_{12}, \qquad \qquad \dot p_2 = (p_2 + p_1) V_{12} -(p_2 + p_3) V_{23}, \nn
\dot q_3 = p_3^2 + V_{31} + V_{23}, \qquad \qquad \dot p_3 = (p_3 + p_2) V_{23} -(p_3 + p_1) V_{31}, \nn
\ee
These equations are not the same as the conventional equations of motion derived from $H$ (viz. $\ddot q_1 = \dot p_1 = V_{31} - V_{12}$ etc.) and the corresponding trajectories  are not the same\footnote{For example, even if we impose the requirement $p_1+p_2+p_3 =0$, the center of mass of the system $x = (q_1+q_2+q_3)/3$ does not remain motionless --- the canonical momenta of the Hamiltonian $I$ do not coincide with the velocities.}, but the dynamics that they describe is similar to the conventional dynamics: oscillations with beats without a trace of blowup (see Fig. \ref{traj-I}). 

\begin{figure}[ht!]
   \begin{center}
 \includegraphics[width=4 in]{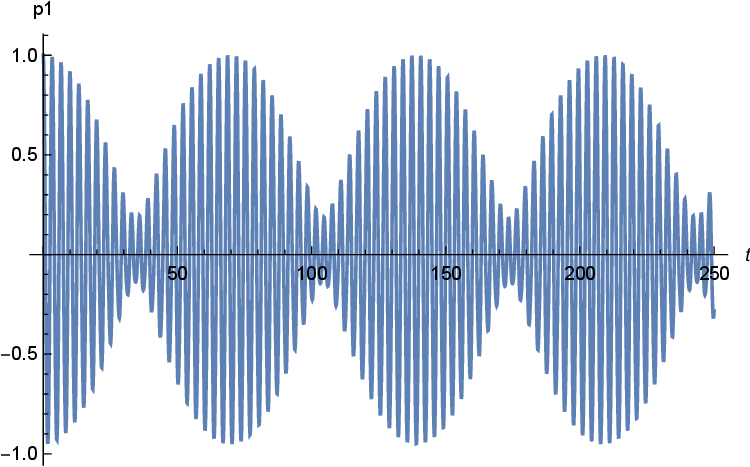}
    \end{center}
\caption{A typical solution to  Eq. \p{eqmot-I} for  $p_1(t)$}
\label{traj-I}
\end{figure} 

One can also perturb the system by adding to $I$ an oscillatory potential:
  \be
\tilde I \ =\ I + \frac \alpha 2 \left[(q_1 - q_2)^2 + (q_1 - q_3)^2 + (q_2 - q_3)^3  \right].
\ee
The Poisson bracket $\{\tilde I, P\}$ still vanishes, but the bracket $\{\tilde I, H\}$ does not. The system is  not integrable now, but the pattern of the trajectories stays the same: oscillations with beats, the only novelty is that the oscillations are not quite regular anymore.

The system \p{Toda} is associated with the algebra $A_2$. In Ref. \cite{Fring-Toda}, Toda systems associated with other algebras like $A_6$, $G_2$ or $B_3$ were studied. It was found that the trajectories derived from Hamiltonians that coincide with one of the higher conserved charges exhibit the same qualitative behavior as in the $A_2$ case.

\end{itemize}

In Ref. \cite{Robert}, both classical and quantum dynamics of the system \p{H-Robert} were studied. In Refs. \cite{Kovner,Pavsic,Deffayet,Damour,Toda}, it was done only for the classical versions. But one can argue that
 
{\it The ghosts in the quantum counterpart of a classical benign ghostly system are benign.}

The argument is simple. If ghosts are not benign at the classical level, the system exhibits a blowup, a collapse in other words. But quantization are associated with quantum fluctuations which make the collapse more difficult. And it may happen that, even though the system collapses classically, it does not happen in quantum case. And if the system is benign classically, no mechanism is seen that can make the quantum system malignant.

\begin{figure}
\bc
\includegraphics[width=.4\textwidth]{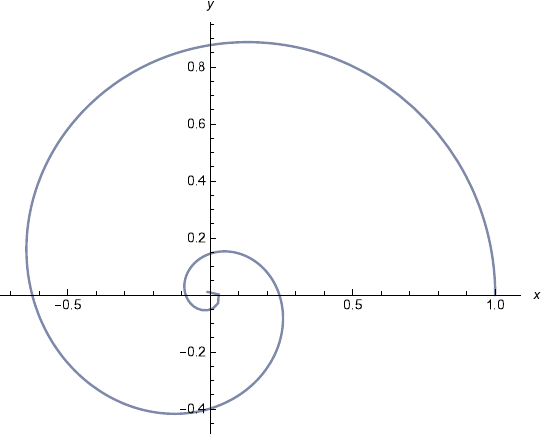}
\ec
        \caption{Falling on the center for the Hamiltonian \p{padenie} with $m = 1$ and
$\kappa = .05$. The energy is slightly negative. The particles with positive energies
escape to infinity.}
\label{fall}
\end{figure}

Probably, the simplest known example of a QM system exhibiting classical and sometimes quantum collapse is the problem of $3D$  motion in a central attractive potential:
 \be
\lb{padenie}
H \  =  \ \frac{\mathbf{p^2}}{2m} - \frac \kappa{r^2}.
 \ee
Classically, for certain initial conditions, the particle falls to the center in a finite time (see Fig. \ref{fall})  The quantum dynamics of this system depends on the value of $\kappa$. 

If $m\kappa < 1/8$, the ground state exists and unitarity is
preserved. And if $m\kappa > 1/8$, the spectrum is not bounded from below and, which is
worse, the quantum problem cannot be well posed until the singularity at the
origin is removed. For example, one can pose $V (r) = -\kappa/r^2$  for $r > a$ and $V (r) =
- \kappa/a^2$  for $ r \leq a$. The spectrum then depends on $a$ \cite{fall}.

Without such a cutoff, probability ``leaks" into the singularity and unitarity is violated. And for $m\kappa \leq 1/8$.
quantum  fluctuations cope successfully with the attractive force of the potential
and prevent the system from collapsing.

This suggests the following general strategy of research. If you wish to find out whether a given quantum ghostly system is benign or not,
there is no need to come to grips with solving Schr\"odinger equation, which might prove to be difficult for a QM system and impossible for a field theory system. It is often sufficient to study  only its  {\it classical} dynamics. If it is benign, one can expect that the quantum dynamics is also benign.\footnote{We are not aware of a rigourous proof of this conjecture, however.}

\section{Ghostly field theories: two dimensions}
\setcounter{equation}0

For me personally, the main motivation to study HD theories has been associated with attempts to build up quantum theory of gravity. For nearly a century, Einstein's general relativity has been successfully resisting all the efforts to quantize it. The action $\int \sqrt{g} R\, d^4x$ describes a nonrenormalizable theory. Besides, it has problems of principal nature, the most acute of them being the loss of causality. It is lost due to the fact that time and spatial coordinates are intertwined in GR and classical equations of motion do not reduce to the Cauchy problem, to evolution in independent flat time.

My dream (it is personal again) is that the Theory of Everything is formulated in a multidimensional flat bulk representing a conventional (supersymmetric?) field theory rather than string theory. And our $4D$ space-time is a brane (a classical solution to the TOE ?) immersed in this flat bulk \cite{dream}. To assure renormalizability, such Holy Grail theory should have dimensionless couplings. And this also means in higher dimensions the presence of higher derivatives.
We will discuss HD gravity in the next section, while here we will dwell on much more simple 2-dimensional HD theories. We will show that in some such theories ghosts are benign.

The first example of a field theory with benign ghosts was suggested in \cite{duhi-pole}. It was a 2-dimensional supersymmetric theory with the Lagrangian representing in the bosonic sector a $2D$ generalization of the Lagrangian \p{L-Robert}:
\be
\lb{duhi-v-pole}
{\cal L} \ =\ \pd_\mu \phi \pd_\mu D - DV'(\phi)
\ee
with 
\be
V(\phi) \ =\ \frac {\omega^2 \phi^2}2 + \frac {\lambda \phi^4}4.
\ee
The model\footnote{By defining $X_\pm = (\phi \pm D)/2$, the kinetic part of the Lagrangian \p{duhi-v-pole} acquires the form
$
(\pd_\mu X_+)^2  - (\pd_\mu X_-)^2.
$
The models with such kinetic term and more general interactions were studied in recent \cite{Deff-Held}.}
 has two integrals of motion: the energy
\be
\lb{E-2D}
E \ =\ \int dx \left[   \dot \phi \dot D + \pd_x \phi \pd_x D + D\phi(\omega^2 + \lambda \phi^2)\right], 
\ee
which can be both positive and negative, and the positive definite
\be
\lb{N-2D}
N \ =\ \int dx \left\{\frac 12 \left[\dot \phi^2 + (\pd_x \phi)^2 \right] + \frac {\omega^2 \phi^2}2 + \frac{\lambda \phi^4}4   \right\}.
 \ee
With only two integrals of motion for the infinite number of degrees of freedom, the model is not integrable and can only be solved numerically.

The classical equations of motion read:
\be
\Box \phi + \omega^2 \phi + \lambda \phi^3 \ =\ 0, \nn
\Box D + D(\omega^2 + 3\lambda \phi^2) \ =\ 0
 \ee
Their solution exhibits an oscillatory behavior. The function $\phi(t,x)$ cannot grow with time: such   growth is incompatible with the conservation of $N$. The
amplitude of the oscillations of the field $D(t,x)$ can grow with time, however.

We played with different values of the parameters $\omega, \lambda$ and with different initial
conditions and never found a collapse, only at worst a linear growth of $D(t,x)$ with
time. Being not integrable, the system exhibits chaotic behavior associated with Lyapunov instability. A typical pattern is displayed  in Fig. \ref{chaos} where the dispersion $d = \sqrt{\langle D^2 \rangle_x}$ is plotted as a function of time. Due to chaoticity, d is not a smooth function of $t$, but exhibits fluctuations. Still the linear
growth trend is clearly seen.

 \begin{figure} 
\begin{center}
\includegraphics[width=0.6\textwidth]{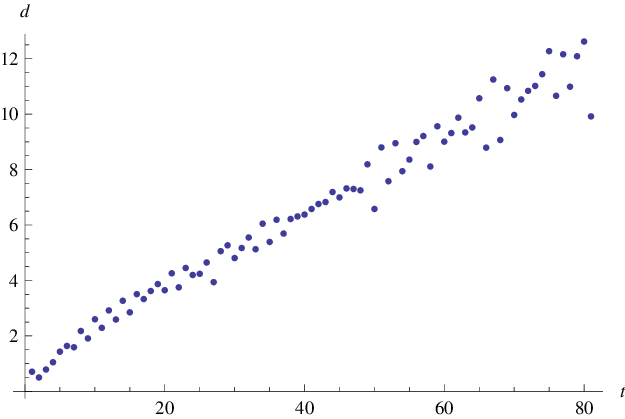}
        \end{center}
    \caption{Dispersion $d = \sqrt{\langle D^2 \rangle_x}$ as a function of time. }
\lb{chaos}
    \end{figure}

\vspace{1mm}

The model \p{duhi-v-pole} is not integrable, but  many integrable 2D field theories including exact soliton solutions are known. An archetypical example of such a theory is the Korteweg--de Vries model.
The ordinary KdV equation for a function $u(t,x)$ reads
 \be
\lb{KdV}
u_{xxx} + 6 u u_x + u_t \ =\ 0.
 \ee
It derives from the action $S = \int dt dx\, {\cal L} [\psi(t,x)]$  with the two-dimensional
Lagrangian density
 \be
\lb{L-KdV}
{\cal L} \ =\ \frac 12 \psi^2_{xx} - \psi_x^3 - \frac 12 \psi_t \psi_x
 \ee
  after having varied over $\psi(t,x)$ and identifying $\psi_x(t,x)$ with $u(t,x)$.

As you see, it looks as a HD theory, but not quite: the Lagrangian involves higher derivatives over $x$, not over $t$. The ordinary Cauchy problem for this theory, when one poses initial conditions for $u(t,x)$ at $t=0$ and follows the time evolution is completely benign and well known. We are interested however in theories involving higher {\it time} derivatives, and the KdV system seems to be of no relevance. What we proposed to do however \cite{Damour} was to {\it rename} the independent variables calling time $t$ the coordinate $X$ and the coordinate $x$ the time $T$.  Then we will be interested in evolution of a genuine HD  system thus obtained in time $T$ (or in old terms, in coordinate $x$). We formulate a ``rotated" Cauchy problem posing the initial conditions for $u(t,0) = u_0(t), u_x(t,0) = u_1(t)$ and $u_{xx}(t,0) = u_2(t)$ with some arbitrary $u_{0,1,2}(t)$   and follow its evolution along $x$. And we pose the same question that we earlier posed for the mechanical systems: is this evolution benign? Does a blowup occur? 

And the answer to this question  is positive. Indeed, the following time-independent function,
 \be
u(t,x) \ =\ - \frac 2 {(x-c)^2}
 \ee
with an arbitrary constant $c$ represents an {\it exact} solution to the KdV equation \p{KdV}. This function has smooth
Cauchy data on the $x=0$ axis, and blows up on the line $x = c$. Thus, an ordinary rotated KdV system has a malignant ghost. 

It was shown in Ref. \cite{Damour} that the situation is much better for the {\it modified} KdV system  with the equation of motion
\be
\lb{mKdV}
u_{xxx} + 12 u^2 \kappa u_x + u_t \ =\ 0.
 \ee
Actually, if $\kappa$ is negative (the so-called {\it defocusing} case), the blowup is still there. But for positive $\kappa$ (focusing case), the blowup is absent. This is a {\it benign} ghostly field theory system.

Well, it is benign according to our definition, but is not unfortunately benign in other meaning: It develops a {\it Lyapunov instability}. 
The KdV system is stable in the ordinary sense, with respect to time evolution. To see that, consider the linearized equation, 
 \be
\lb{linear}
u_{xxx} + u_t \ =\ 0.
 \ee
It can be solved by decomposing the
solution u(x,t) in plane waves $e^{i(\omega t + kx)}$ This yields the
dispersion law, 
 \be
\lb{disp-law}
\omega = k^3.
  \ee
 If one poses the usual Cauchy problem with some Fourier-transformable
initial data,
$$
u(0,x) = v(x) = \int  \frac {dk}{2\pi} v(k) e^{ikx},
$$
the $t$ evolution of the initial data $v(x)$ yields the solution
$$
 u(t,x) \ =\ \int  \frac {dk}{2\pi} v(k) e^{i[\omega(k)t + kx]} = \int  \frac {dk}{2\pi} v(k) e^{i(k^3t + kx)}.
$$
One can show that this solution is smooth.

On the other hand, if one considers the $x$-evolution
Cauchy problem, one starts from three independent
functions, $u_{0,1,2}(t)$ at $x=0$ 
Assuming that the $u_a(t)$ 
are Fourier-transformable, we can represent them as
$$
u_a(t) \ =\ \int \frac {d\omega}{2\pi} u_a(\omega) e^{i\omega t}.
$$
To find how $u(t,x)$ evolves, the dispersion
law \p{disp-law} must now be solved for $k$ in terms of $\omega$. And this equation has
three different {\it complex} roots, namely,
 $$
k_a(\omega) \ =\ \omega^{1/3} z^a
$$
with $z = \frac{-1 + i \sqrt{3}}2$.
This yields a solution for $u(t,x)$ of
the form,
\be
u_a(t) \ =\ \sum_{a=0,1,2} \int \frac {d\omega}{2\pi} v_a(\omega) e^{i[\omega t + k_a(\omega) x] },
\ee
where $v_a(\omega)$ can be expressed via $u_a(\omega)$.

The point of this exercise was to exhibit the fact that, when
considering the $x$ evolution with arbitrary Cauchy data
$u_a(t)$, the solution involves exponentially
growing modes in the $x$ direction, linked to the fact
that  $ik_2 = i \omega^{1/3}z^2$ has a positive real part $\frac{\sqrt{3}}2 \omega^{1/3}$.

Numerical solution of the mKdV equation with different initial conditions on the slice $x=0$ support this heuristic argument. In all the cases studied in \cite{Damour,Praga,Fring-KdV}, a horizon in $x$ was observed beyond which the trajectory is blurred by a high-frequency noise associated with Lyapunov instability.

\begin{figure} [ht!]
      \begin{center}
    \includegraphics[width=.5\textwidth]{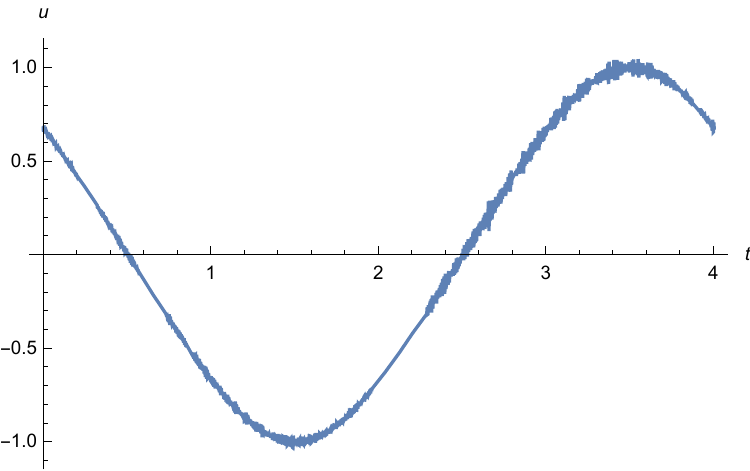}                  
     \end{center}
    \caption{HF noise for the periodic soliton evolution. $x = 4.5$.}        
 \label{noise}
    \end{figure}

This is illustrated in Fig. \ref{noise}. The mKdV equation \p{mKdV} with positive $\kappa$ admits a {\it periodic} soliton solution. It is convenient to set $\kappa =1/2$, in which case the solution reads  \cite{Praga}:
 \be
\lb{sol-cn}
u(t,x) \ =\ {\rm cn}[\sqrt{3}(x+t), m],
 \ee
where cn$(z)$ is the Jacobi elliptic cosine function with the elliptic modulus $m = 1/3$. It is periodic both in $t$ and in $x$ with the period 
 $$
T = L = \frac 4{\sqrt{3}} K\left( \frac 13 \right) \ \approx \ 4.
 $$
We fixed the initial conditions for $x = 0$ and periodic conditions in time as is dictated by \p{sol-cn}
and then numerically solved \p{mKdV}. The numerical solution should reproduce the exact one,
and it does so for $x \leq 4.$ However, at larger values of $x$, a HF noise appears. The result of the
calculation for $x \in (0, 4.5)$ is given in Fig. \ref{noise}. For still larger $x$, the noise blurs the soliton curve completely.

It seems that the Lyapunov instability is a family trait of all ghostly field theory systems, be they benign or not, integrable or not.

\vspace{1mm}

And what happens in the quantum case? Lyapunov instability does not make a quantum problem ill-defined. Take e.g. the Hamiltonian
\be
H \ =\ p_x^2 + p_y^2 + x^2 y^2.
 \ee
Classically, it exhibits a chaotic behavior. There are some closed orbits \cite{Savvidy}, but they are not stable. But the Schr\"odinger equation with such potential can well be solved and the spectrum is discrete\footnote{Cf. the fact that we mentioned earlier: quantization may make a malignant ghost benign.} \cite{Simon}. But the problem here is that, in quantum field theory, we are usually not are interested in the spectrum of the Schr\"odinger equation for the wave functional but rather in scattering amplitudes, the $S$-matrix. To define the latter, we need a well-defined vacuum state and its one-particle excitations. For ghostly theories, neither vacuum nor particle states are  defined.

We do not know today how to treat such theories. But there was time when people did not know how to treat ghostly QM systems, and now we understand that  well enough... 

\section{Quadratic gravity}
\setcounter{equation}0

As was mentioned before, our personal interest to HD theories is fueled by the dream that the Holy Grail Theory of Everything that includes also gravity represents a variant of a HD theory. In this section, we will discuss what is known today about the HD versions of gravity in their classical and quantum aspects. The Einstein's  equations derived from the action 
\be
\lb{S-Ein}
S \ =\ \frac 1{16\pi G} \int \sqrt{-g} R d^4x
  \ee
are of the second order. The action \p{S-Ein} involves the dimensionful Newton's constant, $G = 1/m_P^2$. As a result, the quantum theory is not renormalizable.  On the other hand, the action\footnote{We do not write the term $\propto  \int \sqrt{-g}\, R_{\mu\nu\rho\sigma}R^{\mu\nu\rho\sigma}  d^4x$ because it reduces to \p{S-R2} up to an integral of a total derivative
$$
I_{\rm Gauss-Bonnet} \propto \int \sqrt{-g}\, \left[R_ {\mu\nu\rho\sigma}R^{\mu\nu\rho\sigma} - 4 R_{\mu\nu}  R^{\mu\nu}+ R^2 \right] d^4x $$.}
  \be
\lb{S-R2}
S \ =\ \frac 1{\alpha^2} \int \sqrt{-g} R^2 d^4x + \frac 1{\beta^2} \int \sqrt{-g} R_{\mu\nu} R^{\mu\nu}d^4x
  \ee
includes dimensionless constants $\alpha^2$ and $\beta^2$ and is renormalizable \cite{Stelle}. Moreover, it is {asymptotically free} \cite{Fradkin} \footnote{As is well known, the same behavior is characteristic for the Yang--Mills theory and for QCD. } --- both effective couplings $\alpha(\mu)$ and $\beta(\mu)$ fall down when a characteristic energy scale $\mu$ tends to infinity.
 A particularly nice theory is 
\be
\lb{Weyl} 
S  = \frac 1{\gamma^2}  \int \sqrt{-g}\, C_{\mu\nu\rho\sigma}C^{\mu\nu\rho\sigma}  d^4x  = \frac 2 {\gamma^2}  \int \sqrt{-g}  \left[ R_{\mu\nu} R^{\mu\nu} 
- \frac 13 R^2 \right] d^4x + \ \mbox{int. of a tot. der.},
 \ee
where $C_{\mu\nu\rho\sigma}$ is the Weyl tensor. Besides the general covariance, this action is invariant under the local scale transformations, $g_{\mu\nu}(x) \to \lambda(x)g_{\mu\nu}(x)$. However, loops bring about conformal anomaly, and even if conformaly not invariant structure $R$ were absent in the tree-level Lagrangian, it would be  generated at the one-loop level.

Conformal anomaly in an asymptotically free theory brings about a mass scale, which was absent in the tree action \p{S-R2}. This phenomenon of {\it dimensional transmutation} is well known in Yang--Mills theory. There also the coupling constant in the tree Lagrangian carries no dimension\footnote{This applies also to QCD with massless quarks.}, but the physical mass scale $\Lambda_{YM}$ (the confinement scale) is nevertheless generated. It is a scale where the coupling constant, which falls down when a characteristic energy increases but  goes up when the energy decreases, reaches the values of order 1. The effective Lagrangian which describes the hadron dynamics at low energies does not look similar to ${\cal L}_{QCD}$. It is Yukawa Lagrangian, effective chiral Lagrangian, etc. 

And the same happens to quadratic gravity \cite{spontan}. A mass scale (associated with the Planck scale $m_P$), where the coupling constants $\alpha, \beta$ reach the values of order 1, is generated. The effective action is still generally covariant. It must thus involve a huge cosmological term,\footnote{Experimentally, it is very small; nobody understands why.}
$ \sim \  m_P^4 \int \sqrt{-g} \,d^4 x$
and also the Einstein term$ \sim m_P^2\int \sqrt{-g}\, R \, d^4 x  $,

In fact, this mechanism is a variant of the old Sakharov mechanism \cite{Sakharov}. Sakharov noticed that the induced Einstein--Hilbert term in the effective action  is generated together with cosmological term in {\it any} theory including a mass parameter. It appears e.g. in QED --- consider a vacuum electron loop in a curved background. A specifics of quadratic gravity is that the mass parameter is not explicitly present in the Lagrangian but appears as a result of dimensional transmutation. 

The quadratic gravity including or not  the Einstein--Hilbert term is a HD theory. Therefore it includes ghosts. The question is whether these ghosts are benign or malignant. Well, dealing with  gravity, we cannot really study a quantum problem: at the present level of understanding, quantum gravity is not a well-defined theory. But one can study it at the classical level. According to Definition \ref{ben-clas-ghost}, a benign ghostly classical system is a system not involving collapsing trajectories. But {\it any} gravity theory including Einstein's theory without quadratic terms involves the phenomenon of gravitational collapse with formation of black hole horizon and the singularity inside. Does it mean that all the gravity theories are inherently malignant? Not necessarily. One can recall that in physics only those phenomena have sense that can be tested in experiment. And formation of horizon takes an infinite time on the clocks of a distant observer, and the horizon (not speaking of its interior) can never be observed.

The collapse of dust into a black hole was recently studied for the theory with the action including  the terms $ R$ and $R^2$ in Ref. \cite{Fante}. They used the spherically symmetric ansatz where the Weyl tensor vanished and did not include in the action by that reason the term $\propto C_{\mu\nu\rho\sigma}C^{\mu\nu\rho\sigma}$. They found that the dynamics of the collapse is similar to what happens in general relativity. As is the case for Einstein's theory, formation of horizon takes an infinite time for an external observer. It was also found that the presence of $R^2$ term makes the collapse a little faster at the final stage.

What has not been done so far is a study of the generic (3+1) problem without assuming spherical symmetry.  In this case, the Weyl-squared term should also be included in the action. One can expect that the spherically symmetric collapse scenario would be unstable in Lyapunov sense as it is in pure Einstein's theory \cite{chaos-grav}. But whatever the solution of this interesting problem is, we do not think that the system can run into singularity in finite time for distant observer. In gravity, the singularities are protected by horizons.\footnote{This is the so-called {\it cosmic censorship} hypothesis \cite{censor}. It is not proven, but looks very plausible.} And the formation of the horizon (when the signals from the collapsing matter cease to reach the distant observer) must take an  infinite time. 

{\it To conclude}: quadratic gravity is probably benign in the same sense as simple $2D$ models considered in the previous section are: a blowup is absent, but Lyapunov instability is present.

\section{Supersymmetric ghostly models}
\setcounter{equation} 0

As was mentioned, the model \p{L-Robert} describes in fact a bosonic sector of a certain supersymmetric mechanical model. The latter includes besides $x$ also   a Grassmann variable $\psi$. The full supersymmetric Lagrangian\footnote{It can be derived using superfield formalism, but  we will quote only its component expression.} reads
\be
\lb{super-Robert}
L \ =\  \dot x \dot D  + \dot{\bar \psi} \dot \psi - V'(x) D - V''(x) \bar \psi \psi .
\ee
The $2D$ version \p{duhi-v-pole} of \p{L-Robert} describes a bosonic sector of a certain $2D$ supersymmetric field model. The full Lagrangian is a straightforward generalization of \p{super-Robert}:
\be
\lb{super-2D}
{\cal L} \ =\  \pd_\mu \phi \pd_\mu D  + \pd_\mu \bar \psi \pd_\mu  \psi - V'(\phi) D - V''(\phi) \bar \psi \psi,  \qquad \qquad \mu = 0,1.
\ee

\vspace{1mm}

Supersymmetric extentions of  quadratic gravity models also exist. There are ${\cal N} = 1$, ${\cal N} = 2$, ${\cal N} = 3$ and ${\cal N} = 4$ extensions. The ${\cal N} = 4$ extension of the conformal gravity \p{Weyl} is expecially interesting. It turns out that the $\beta$ function in this theory {\it vanishes} in all orders \cite{super-Fradkin} ! As a result, there are no ultraviolet divergences, the coupling constant does not run and the superconformal symmetry of the classical Lagrangian is also a symmetry of quantum theory. This theory includes besides the metric the following bosonic fields:
\begin{itemize}
\item complex symmetric    scalars $E_{(ij)}$, \ $i,j = 1,2,3,4$,
\item complex scalars $D^{[ij]}_{[kl]}$ satisfying the constraints $D^{ij}_{kj} =0$, 
\item vector gauge field $V_{j,\mu}^i \in su(4)$,
\item tensor $T^{ij}_{\mu\nu}$ which is antisymmetric in $i \leftrightarrow j$ and antiself-dual in the indices $\mu,\nu$. 
 \end{itemize}
The fields $E,V,T$ carry canonical dimension 1 and the field $D$  canonical dimension 2.  In addition, there is a scalar $\phi$ of canonical dimension 0.

There is also fermion fields, which are of no relevance for our purposes. The full number of bosonic  on-shell degrees of freedom coincides of course with the full number of fermionic on-shell degrees of freedom, as is the case in any supersymmetric theory.
 
The bosonic part of the action reads\footnote{See Eq. (3.30) in Ref. \cite{super-Fradkin}, which we've polished a bit and removed irrelevant for us numerical coefficients.}
\be
\lb{Lbos-superconf}
&&S_{\rm bos} \   = \  \frac 1{\gamma^2} \int  {\sqrt{-g}} \left[ C_{\mu\nu\rho\sigma}C^{\mu\nu\rho\sigma}  + F_{j,\mu\nu}^i(V)   
F_{i}^{j,\mu\nu}(V)   +  \varepsilon^{ijkl} \,{\cal D}^\mu T_{ij,\mu\nu} \, {\cal D}_\rho T^{*\rho\nu}_{kl} \right. \nn
&& \qquad  \qquad  -   \varepsilon^{ijkl} {\cal D}^\mu E_{ij} {\cal D}_\mu E^*_{kl} + {\cal D}^2 \phi  {\cal D}^2 \phi^*  +   \varepsilon^{ijkl} \varepsilon_{mnpq}\left. D_{ij}^{mn} D_{kl}^{*pq}\right]\, d^4x,
 \ee
where $F_{j,\mu\nu} ^i$ are the field densities of the gauge field $V_{i,\mu}^j$ and ${\cal D}_\mu$ are the gravitational covariant derivatives.
We see that the field $D^{ij}_{kl}$ enters only with the term $\sim D^2$. The corresponding equations of motion are $D^{ij}_{kl}=0$, so that this field does not affect the dynamics and can be eliminated.

We believe that the ghosts in this theory are benign in the same sense as in the ordinary nonsupersymmetric quadratic gravity. Singularities are protected by horizons, which a system cannot reach in a finite time by the clocks of a distant observer. By the same token as in pure quadratic gravity, the collapsing trajectories are probably not stable in the Lyapunov sense.

The terms in the Lagrangian involving the fields $E,T$ and $V_\mu$ have a conventional form not bringing about ghostly dynamics. The same concerns the term ${\cal D}^2 \phi {\cal D}^2 \phi^*$ --- the corresponding equation of motion ${\cal D}^4 \phi = 0$ does not lead to a blowup in flat space, and we do not see by what mechanism can it arise when space is curved. But these are only plausible arguments and it would be interesting to confirm them by a numerical analysis.

\vspace{1mm}

The final model that we wish to mention here is  a higher-derivative supersymmetric gauge model in 6 dimensions. This model was built up in Ref. \cite{Zupnik}. It involves higher derivatives. Again, we will not discuss here the beautiful details of its supersymmetric construction and only present the action in the bosonic sector. It involves a $6D$ gauge field $A_M \in \mathfrak{g}$ ($M = 0,1,2,3,4,5$) and the scalars $D_{jk}$ ($j,k = 1,2$) also belonging to the gauge algebra $\mathfrak{g}$ and carrying canonical dimension 2. The action has a rather simple form,
\be
\lb{gauge-HD}
S \ =\ \frac 1{g^2} \int  d^6x  {\rm Tr} \left\{ (\nabla^M F_{MN})^2 + \frac 12 (\nabla_M D_{jk})^2   + D_{lk} D^{kj} D_j\,^l   \right\}
+ {\rm ferm. \ terms}
\ee
with $\nabla_M = \pd_M - iA_M$ and the indices $j,k = 1,2$ being raised and lowered by $\varepsilon^{jk} = - \varepsilon_{jk}$.

This is a higher-derivative theory. In contrast to the ordinary $6D$ gauge theory with the action $S  = {\cal M}^2 \int d^6x \, {\rm Tr} \left\{  F_{MN}^2 \right\} + \ldots$, which involves a dimensional coupling constant ${\cal M}$ and is not renormalizable, the coupling constant $g^2$ in the action \p{gauge-HD} carries no dimension and the theory is renormalizable. The effective coupling constant runs here. One can derive \cite{Zupnik, Casarin}:
  \be
  \lb{beta-6D}
g^2(\mu) \ =\ g_0^2 \left(1 - \frac {11 \,g_0^2 c_V}{96 \,\pi^3} \ln \frac \Lambda \mu \right),
  \ee
where $c_V$ is the adjoint Casimir eigenvalue [$c_V = N$ for $SU(N)$], $\mu$ is a characteristic energy scale and $g_0$ is the constant defined at the ultraviolet scale $\Lambda$. We see that the coupling grows with energy. This is the Landau zero situation, as in the ordinary 4-dimensional QED.

One could also flip the sign in  the definition of the action \p{gauge-HD} \cite{Buch}. In ordinary theories like QED or QCD, the sign is fixed by the requirement for the energy to be positive. But here we are dealing with a higher-derivative ghostly theory where the energy can be both positive and negative, and both signs of $g^2$ are allowed. If we flipped the sign in \p{gauge-HD}, the absolute value of $g^2$ would decrease with energy and we would arrive at the situation of asymptotic freedom.

Are the ghosts here benign or not? Unfortunately, {\it not}. This is seen if we restrict our attention to the dynamics of the fields $D_{jk}$ in 
\p{gauge-HD}. These fields have many components, but their dynamics is roughly the same as the dynamics of the field $\phi$ in the 6-dimensional model, $${\cal L} = \frac 12 (\pd_\mu \phi)^2 + \lambda \phi^3.$$
The equation of motion is
 \be
\lb{eqmot-phi}
\Box \phi - 3\lambda \phi^2 \ =\ 0.
 \ee
The homogeneous field 
\be
\phi(t, \vecg{x}) = \frac 2{3\lambda(t-t_0)}
 \ee
is a solution to this equation. At $t = t_0$, it becomes singular.

 \section{Conclusions}

Higher-derivative theories attracted attention of scholars since the times of Euler and Ostrogradsky. They have also attracted a considerable attention recently, and today we understand their dynamics much better than we did 20 years ago. The HD systems have ghostly nature: 
their energy is not bounded neither from below nor from above and can acquire any real value. In their quantum versions, the spectrum of the Hamiltonian does not have a bottom (a ground state). Neither has it a ceiling.  Many such theories are pathological: their classical trajectories are plagued by singularities and their quantum versions are plagued by loss of unitarity. 

But there exist also benign HD theories where this does not happen. Benign mechanical HD systems often enjoy Lyapunov stability, but the HD field theories that have been studied so far do not have this property: when a generic initial condition is chosen, the evolution of the system exhibits chaotic behavior.

The author of this text believes that  the Holy Grail fundamental Theory of Everything  includes higher derivatives in its formulation.


\begin{thebibliography}{96}

\bibitem{Euler} L. Euler, {\it Methodus inveniendi lineas curvas maximi minimive proprietate gaudentes,
sive solutio problematis isoperimetrici latissimo sensu accepti}, Lausanne and
Geneva: Bousquet, 1744;

C. Fraiser, {\it Leonhard Euler, book on the calculus of variations (1744)}, in: [Landmark writings in western mathematics 1640--1940, Elsevier], pp. 168-180.

\bibitem{brachist} See the article {\it Brachistochrone curve} in Wikipedia. 

\bibitem{Ostrolater}  Istoriko-Matematicheskie Issledovania {\bf XVI} (1965) 49.

\bibitem{Ostroflux} M. Ostrogradsky, {\it Note sur la th\'eorie de la chaleur}, Mem. Acad. St. Petersbourg {\bf I} (1831)  29.

\bibitem{Katz} V.J. Katz, {\it The history of Stokes' theorem}, Math. Magazine {\bf 52}  (1979) 146. 

\bibitem{OstroHam} M. Ostrogradsky, {\it M\'emoire sur les \'equations diff\'erentielles relatives au probl\`eme des isop\'erim\`etres}, Mem. Acad. St. Petersbourg {\bf VI}  (1850) 385.

\bibitem{Pagani} E. Pagani, G. Techhiolli and S. Zerbini, {\it On the problem of stability for higher-order derivative Lagrangian systems}, Lett. Math. Phys. {\bf 14} (1987) 311.

\bibitem{Eliezer} D.A. Eliezer and R.P. Woodard, {\it The problem of nonlocality in string theory}, Nucl. Phys. {\bf B325} (1989) 389. 

\bibitem{Woodard} R.P. Woodard, {\it Ostrogradsky's theorem on Hamiltonian instability}, Scholarpedia {\bf 10} (2015) 32243.

\bibitem{Podolsky} B. Podolsky and P. Schweb, {\it Review of a generalized electrodynamics}, Rev. Mod. Phys. {\bf 20} (1948) 40.

\bibitem{PU} A. Pais and G.E. Uhlenbeck, {\it On field theories with non-localised action}, Phys. Rev. {\bf 79} (1950) 145.

\bibitem{Mannheim} P.D. Mannheim and A. Davidson, {\it Dirac quantization of the Pais--Uhlenbeck fourth order oscillator}, Phys. Rev. {\bf A71} (2005) 042110.

\bibitem{Bolonek} K. Bolonek and P. Kosinski, {\it Comments on``Dirac quantization of the Pais--Uhlenbeck
fourth order oscillator"}, arXiv:quant-ph/0612009.

\bibitem{Singapur} A. Smilga, {\it Classical and quantum dynamics of higher-derivative systems}, IJMP {\bf A32} (2017) 1730025.

\bibitem{eston} M. Raidal and H. Veerm\"ae, {\it On the quantisation of complex higher derivative theories
and avoiding the Ostrogradsky ghost},     Nucl.Phys. {\bf B916} (2017) 607.

\bibitem{Kovner} I.B. Ilhan and A. Kovner, {\it Some comments on ghosts and unitarity: the Pais-Uhlenbeck oscillator revisited}, Phys. Rev. {\bf D88} (2013) 044045.

\bibitem{Bender} C. Bender and P. Mannheim, {\it No-ghost theorem for the fourth-order derivative Pais-Uhlenbeck oscillator model}, Phys. Rev. Lett. {\bf 100} (2008) 110402.

\bibitem{LeeWick} T.D. Lee and G.C. Wick, {\it Finite theory of quantum electrodynamics}, Phys. Rev. {\bf D2} (1970) 1033.

\bibitem{Carroll} S.M. Carroll, M. Hoffman and M. Trodden, {\it Can the dark energy equation-of-state parameter $w$ be less than $-1$ ?}, Phys. Rev. {\bf D68} (2003) 023509.

\bibitem{benign} A.V. Smilga, {\it Benign vs. malicious ghosts in higher-derivative theories}, Nucl. Phys. {\bf B206} (2005) 598.

\bibitem{Robert} D. Robert and A.V. Smilga, {\it Supersymmetry vs. ghosts}, J. Math. Phys. {\bf 49} (2008) 042104.

\bibitem{Perelomov} A.M. Perelomov, {\it Integrable Systems of Classical Mechanics and Lie Algebras}, Birkh\"auiser Verlag, 1990.

\bibitem{Pavsic} M. Pav\v{s}i\v{c}, {\it Stable self-interacting Pais--Uhlenbeck oscillator}, Mod. Phys. Lett. {\bf A28} (2013) 1350165;

{\it Pais–Uhlenbeck oscillator and negative energies},  Int. J. Geom. Meth. Mod. Phys. {\bf 13} (2016)  1630015.

\bibitem{Deffayet} C. Deffayet, S. Mukohyama and A. Vikman, {\it Ghosts without runaway}, Phys. Rev. Lett. {\bf 128} (2022) 041301.

\bibitem{Damour} T. Damour and A. Smilga, {\it Dynamic systems with benign ghosts}, Phys. Rev. {\bf D105} (2022) 045018.

\bibitem{Toda} A.V. Smilga, {\it On exactly solvable ghost-ridden systems}, Phys. Lett.
{\bf A389} (2021) 127104.

\bibitem{Fring-Toda} A. Fring and B. Turner, {\it Higher derivative Hamiltonians with benign ghosts from affine Toda lattices}, J. Phys. {\bf A56} (2023) 295 203.

\bibitem{fall} K.M. Case, {\it Singular potentials}, Phys. Rev. {\bf 80} (1950) 797;

K. Meetz, {\it Singular potentials in nonrelativistic quantum mechanics}, Nuovo Cimento {\bf 34}  (1964) 690;

A. M. Perelomov and V. S. Popov, {\it Collapse onto scattering centre in quantum mechanics}, Teor. Mat. Fiz. {\bf 4} (1970) 48.

\bibitem{dream} A.V. Smilga, {\it Quantum gravity as Escher's dragon}, Phys. Atom. Nucl.
 {\bf 66} (2003) 2092;

{\it $6D$  superconformal theory as the Theory of Everything}, in Gribov Memorial Volume, eds. Yu. L. Dokshitzer, P. Levai and Y. Nyiri (World Scientific, 2006), p. 443, arXiv:hep-th/0509022.

\bibitem{duhi-pole} A.V. Smilga, {\it Supersymmetric field theory with benign ghosts}, J. Phys. {\bf A47} (2014) 052001.

\bibitem{Deff-Held}  C. Deffayet, A. Held, S. Mukoyama and A. Vikman, {\it Ghostly interactions in (1+1) dimensional classical field theory},
Phys. Rev. {\bf D112} (2025) 065011.

\bibitem{Praga} A.V. Smilga, {\it Modified Korteweg--de Vries equation as a system with benign ghosts},  Acta Polytechnica {\bf 62} (2022) 190. 

\bibitem{Fring-KdV}  A. Fring, T. Taira and B. Turner, {\it Higher time-derivative theories from space-time interchanged integrable field theories}, Universe {\bf 10} (2024) 198. 

\bibitem{Savvidy} S.G. Matinyan, G.K. Savvidy and N.G. Ter-Arutunian Savvidy, {\it Classical
Yang–Mills mechanics. Nonlinear color oscillations}, Sov. Phys. JETP {\bf 53}
(1981) 421; 

 T.S. Biro, S.G. Matinian and B. Muller, {\it Chaos and Gauge Field Theory}, World
Scientific, Singapore, 1994.

\bibitem{Simon} B. Simon, {\it Some quantum operators with discrete spectrum but classically continuous spectrum}, Ann. Phys. {\bf 146} (1983) 209.

\bibitem{Stelle} K.S. Stelle, {\it Renormalization of higher-derivative quantum gravity}, Phys. Rev. {\bf D16} (1977) 953.

\bibitem{Fradkin} E.S. Fradkin and A.A. Tseytlin, {\it Renormalizable asymptotically free quantum theory of gravity}, Nucl. Phys. {\bf B201} (1982) 469;

I.G. Avramidy and A.O. Barvinsky, {\it Asymptotic freedom in higher-derivative quantum gravity}, Phys. Lett. {\bf B159}, 269.



\bibitem{spontan} S.L. Adler, {\it Einstein gravity as a symmetry-breaking effect in quantum field theory}, Rev. Mod. Phys. {\bf 54} (1982) 729;

A.V. Smilga, {\it Spontaneous generation of the Newton constant in the renormalizable gravity theory}, Preprint ITEP-63 (1982) published in the Proceedings of the conference in Group Theoretic methods in Physics (Zvenigorod, 1982), vol. 2, p. 73, reprinted in arXiv:1406.5613 [hep-th].

\bibitem{Sakharov} A.D. Sakharov, {\it Vacuum fluctuations in curved space and the theory of gravitation}, Sov. Phys. Dokl. {\bf 12} (1968) 1040.

\bibitem{Fante} L. Buonifante, F. di Filippo, I. Kol\`a\v{r} and F. Saueressig, {\it Dust collapse and horizon formation in quadratic gravity}, JCAP, {\bf 01} (2025) 114.

\bibitem{chaos-grav} J.D. Barrow, {\it Chaotic behavior in general relativity}, Phys. Rept. {\bf 85} (1982) 1;

P.S. Joshi and D. Malafarina, {\it Instability of black hole formation in gravitational collapse}, Phys. Rev. {\bf D83} (2011) 024009.

\bibitem{censor} R. Penrose, {\it Gravitational collapse: the role of general relativity}, Rivista del Nuovo Cimento, Numero speziale {\bf 1} (1969) 252.

\bibitem{super-Fradkin} E.S. Fradkin and A.A. Tseytlin, {\it  Conformal anomaly in Weyl theory and anomaly-free superconformal theories}, Phys. Lett. {\bf B134} (1984) 187; 
 {\it Conformal supergravity}, Phys. Repts. {\bf 119} (1985) 233.

\bibitem{Zupnik} E.A. Ivanov, A.V. Smilga and B. Zupnik, {\it Renormalizable supersymmetric gauge theory in six dimensions}, Nucl. Phys. {\bf B726} (2005) 131.

\bibitem{Casarin} L. Casarin and A.A. Tseytlin, {\it One-loop $\beta$ functions in 4-derivative gauge theory in 6 dimensions}, JHEP {\bf 08} (2019) 159.

\bibitem{Buch} I.L. Buchbinder, E.A. Ivanov, B.S. Merzlikin and K.V. Stepanyantz, {\it The renormalization structure of $6D$ ${\cal N} = (1,0)$ supersymmetric higher-derivative theory}, Nucl. Phys. {\bf B961} (2020) 115249. 










\end{thebibliography}
\end{document}